\documentclass[conference]{IEEEtran}
\IEEEoverridecommandlockouts
\pdfoutput=1

\usepackage{amsmath,amssymb,amsfonts}
\usepackage{algorithmic}
\usepackage{graphicx}
\usepackage{textcomp}
\usepackage{xcolor}
\usepackage[utf8]{inputenc} 
\usepackage[T1]{fontenc}    
\usepackage{hyperref}       
\usepackage{url}            
\newcommand*{\refname}{References}
\def\BibTeX{{\rm B\kern-.05em{\sc i\kern-.025em b}\kern-.08em
    T\kern-.1667em\lower.7ex\hbox{E}\kern-.125emX}}
\begin{document}

\title{GenPluSSS: A Genetic Algorithm Based Plugin for Measured Subsurface Scattering Representation\\
}
\author{\IEEEauthorblockN{Barış Yıldırım}
\IEEEauthorblockA{\textit{\.{I}zmir Innovation and Technology, Inc.} \\
baris.yildirim@izmirteknoloji.com.tr}
\and
\IEEEauthorblockN{Murat Kurt}
\IEEEauthorblockA{\textit{International Computer Institute, Ege University} \\
murat.kurt@ege.edu.tr}
}

\maketitle

\begin{abstract}
This paper presents a plugin that adds a representation of homogeneous and heterogeneous, optically thick, translucent materials on the Blender 3D modeling tool. The working principle of this plugin is based on a combination of Genetic Algorithm (GA) and Singular Value Decomposition (SVD)-based subsurface scattering method (GenSSS). The proposed plugin has been implemented using Mitsuba renderer, which is an open source rendering software. The proposed plugin has been validated on measured subsurface scattering data. It's shown that the proposed plugin visualizes homogeneous and heterogeneous subsurface scattering effects, accurately, compactly and computationally efficiently.
\end{abstract}


\section{Introduction}
Optically thick, translucent materials exhibit complex light scattering behaviors, as they include subsurface scattering effects. Human skin, wax and marble are examples of translucent materials. Light scattering occurs at various layers at the material and this situation provides some of this inner part of material to be seen. In the field of computer graphics, realistic representation of optically thick, translucent materials requires modeling of the Bidirectional Surface Scattering Reflectance Distribution Function (BSSRDF), which is a generalization of the Bidirectional Reflectance Distribution Function (BRDF) introduced by Nicodemus et al.~\cite{Nicodemus77},~\cite{Kurt2009SIGGRAPHCG,Kurt2018DEU}. 

Translucent materials can be decomposed into two classes. The first one is homogeneous materials and the other one is heterogeneous materials. In computer graphics, accurate measurement and representation of heterogeneous translucent materials is one of the challenging problems, as it requires to handle large data size, and modeling heterogeneous translucent materials is more complicated than representing homogeneous translucent materials~\cite{Frisvad20}.

In this study, we use genetic optimization and SVD-based subsurface scattering approach (GenSSS) for compact and accurate representation of heterogeneous, optically thick, translucent materials. This approach was implemented in Mitsuba renderer~\cite{Mitsuba} and included as an integration plugin into the Blender 3D modeling tool~\cite{Blender}. Our plugin (GenPluSSS) provides to visualize homogeneous and heterogeneous translucent materials accurately, and computationally efficiently. As it's seen in Figure~\ref{fig:hetorenders}, heterogeneous subsurface scattering effects can be demonstrated visually plausibly by using the proposed plugin.

In summary, the main contributions of this paper are:
\begin{itemize}
    \item A novel plugin for representing both homogeneous and heterogeneous optically thick translucent materials.
    \item A tuneable plugin to trade visual complexity against rendering times.
    \item A detailed validation of our plugin (GenPluSSS).
    \item A comparison to the state-of-the-art, showing significant improvements in terms of rendering times of homogeneous translucent materials.
\end{itemize}

\section{Related Work}

Our proposed plugin is based on Kurt's~\cite{Kurt2020MAM, Kurt2021TVC} GenSSS model, which builds upon compression based representations, GAs, and BSSRDF representations. It's also related to plugins for BSSRDF representations, so we briefly review each of these below.

\noindent{\textbf{BSSRDF representations:}} Jensen et al.~\cite{Jensen01} presented the diffusion dipole approximation for homogeneous subsurface scattering representation. This approach is effective on representing homogeneous translucent materials. The main disadvantage of this approach is that it can't represent heterogeneous translucent materials accurately. Jensen and Buhler~\cite{Jensen02} suggested a two-step hierarchical algorithm to calculate the dipole diffusion approximation. This algorithm is faster than Jensen et al.'s~\cite{Jensen01} BSSRDF approximation. d'Eon et al.~\cite{dEonetal07} suggested a multi-layered subsurface scattering representation for real-time rendering of human skin. Jimenez et al.~\cite{Jimenez09} extended Jensen et al.’s BSSRDF approximation to represent human skin in real time. To represent human skin, the diffusion dipole approximation was extended by diffusion multipole approximation, which was proposed by Donner and Jensen~\cite{Donner05}. Jakob et al.~\cite{Jakobetal10} extended Jensen et al.’s BSSRDF approximation by using an anisotropic approach. Yatagawa et al.~\cite{yatagawa2020linsss} suggested a real-time screen-space rendering technique for representing the heterogeneity of subsurface scattering. The proposed technique is called as LinSSS. However, their goal is not to represent smaller approximation errors, but to provide an approximation that can be used in real-time screen-space rendering. 

\noindent{\textbf{Compression based representations:}} In computer graphics, factorization-based representations have been used for representing BRDFs~\cite{Sun07,Bigili2011CGF,Tongbuasirilai2020TVC}, BTFs~\cite{RuitersK09,RuitersSK12}, homogeneous subsurface scattering~\cite{Jimenez15}, and heterogeneous subsurface scattering~\cite{Peers06,Kurt2013EGSR,Kurt2013TPCG}. Yatagawa et al.~\cite{yatagawa2020data} suggested a data compression method for measured heterogeneous subsurface scattering, since the large data sizes cause problems for devices, which have limited memory, such as tablets and mobile phones in real-time rendering.

\noindent{\textbf{Genetic algorithms:}} Genetic algorithms are less applied to problems in computer graphics and have been used extensively in optimization problems~\cite{Kurt2020MAM, Kurt2021TVC}. Kurt~\cite{Kurt2020MAM, Kurt2021TVC} used GA and reported that when compared to inverse-rendering-based techniques~\cite{Masia09,Munoz09}, his GA is computationally efficient, as his fitness function is computed between measured and factored data for each candidate chromosome. 

\noindent{\textbf{Plugins for BSSRDF representations:}} Blender 3D modeling tool~\cite{Blender} includes a plugin for the diffusion dipole approximation, which is used to represent optically thick, homogeneous translucent materials. Styperek and Juh\'e~\cite{BlenderPlugin} proposed the plugin, which was developed by using Mitsuba renderer~\cite{Mitsuba}. \"{O}nel et al.~\cite{Onel2014EG,Onel2019PL} proposed a plugin for representing optically thick, heterogeneous translucent materials. The proposed plugin is based on an SVD approach proposed by Kurt~\cite{Kurt2014PhDThesis}. Similar to Styperek and Juh\'e~\cite{BlenderPlugin}, \"{O}nel et al.~\cite{Onel2014EG,Onel2019PL} prepared the plugin for Blender 3D modeling tool by using Mitsuba renderer~\cite{Mitsuba}.    

Unlike previous works, Kurt's GA~\cite{Kurt2020MAM, Kurt2021TVC} combined with the SVD-based technique provides a new approach for representing measured subsurface scattering data of translucent materials. This was the main motivation of our work. His algorithm was imported via our plugin (GenPluSSS) after his work. Thus, homogeneous and heterogeneous subsurface scattering representations was made for Blender 3D modeling tool.

\section{Our Integration Plugin (GenPluSSS)}
\begin{figure}[t]
    \centering
    \includegraphics[width=0.94\linewidth]{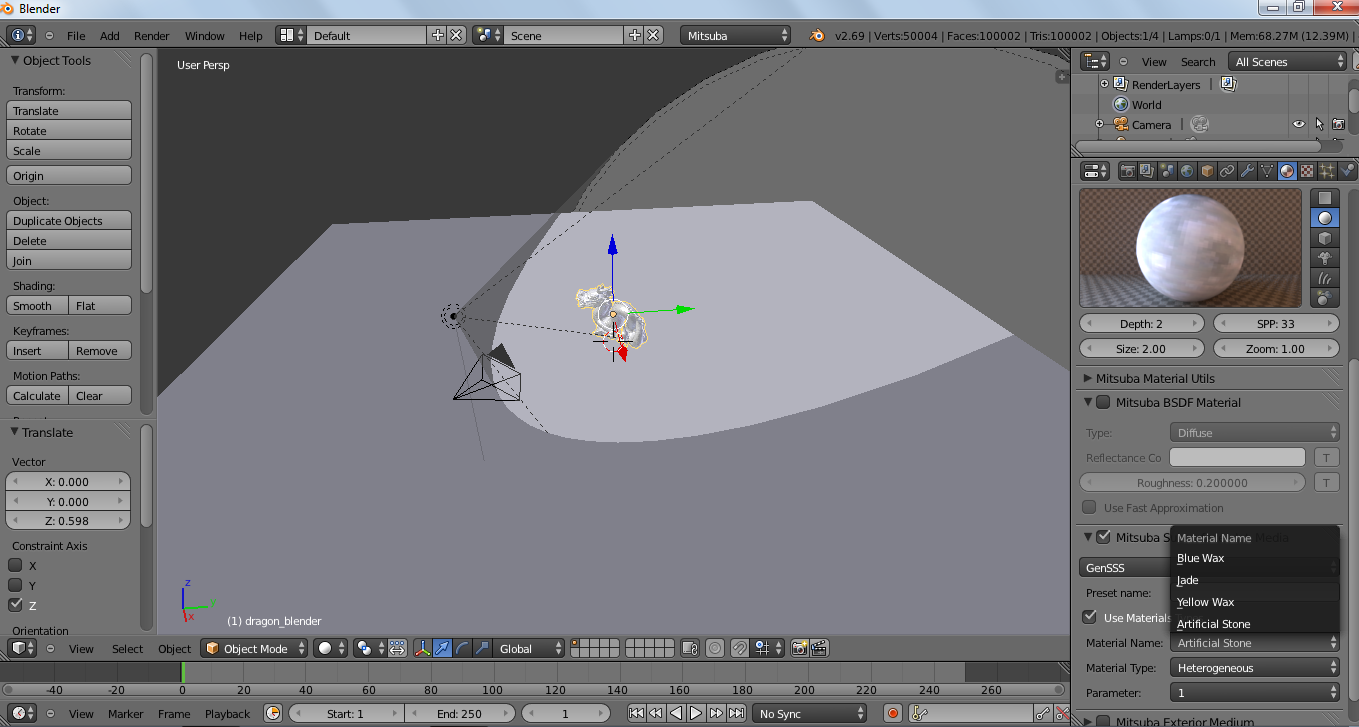}
    \caption{A general overview of the Blender 3D Modeling Tool [12] and our integration plugin (GenPluSSS) with preview scene of Heterogeneous Artificial Stone material was represented. The value of \textit{K} parameter in GenSSS model was selected as 1.}
    \label{fig:blenderview}
\end{figure}

In this work, our goal is to develop a plugin for rendering optically thick, translucent materials. The proposed plugin can be considered as an interface between the renderer and the software modeling tool. Properties of lighting, shading, texture mapping in the scene are defined by the 3D modeling tool. Then, to get final image, the scene configuration is sent to the renderer. Light sources, materials and their properties can be easily configured by using the 3D modeling tool. The Blender~\cite{Blender} is an open source developing platform. The Blender uses Python~\cite{Python} as a scripting language, and we can call C++ functions easily by using Python. As Mitsuba renderer~\cite{Mitsuba} has been written by C++ programming language, and an open source renderer, we opted to choose the Blender 3D modeling tool and Mitsuba renderer to prepare our integration plugin (GenPluSSS). Other advantages of selecting Mitsuba~\cite{Mitsuba} as an open source renderer are performance, scalability, durability, reality, accuracy and ease of use. The proposed plugin was written by using Python programming language, so interaction between our plugin (GenPluSSS) and Mitsuba become much more easier. In this study, we used Mitsuba version of 0.5.0 and Blender version of 2.69. A general overview of the Blender 3D modeling tool and our integration plugin (GenPluSSS) can be seen in Figure~\ref{fig:blenderview}.     

\begin{figure}[t]
    \centering
    \includegraphics[width=0.94\linewidth]{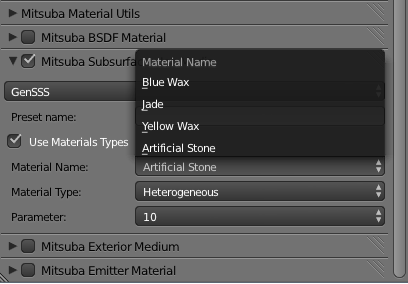}
    \caption{ The graphical user interface of our integration plugin (GenPluSSS) in the Blender 3D Modeling Tool [12].}
    \label{fig:selectscreen}
    \vspace{-0.4cm}
\end{figure}

Material name, material type and parameter selection can be made by using our proposed plugin. In this study, as we used measured data sets from Song et al.~\cite{Song09}, material names include "Artificial Stone", "Blue Wax", "Jade", and "Yellow Wax", which can be seen in Figure~\ref{fig:selectscreen}. In addition, "Homogeneous" or "Heterogeneous" material types can be selected by the material type (see Figure~\ref{fig:selectscreen}). If the material type is selected as heterogeneous, the value of \textit{K} parameter -which's name is "Parameter" in the plugin- can be determined by the user as "1", "5" or "10". However, if the material type is selected as Homogeneous, the value of \textit{K} parameter selection will be invisible and in this case the default parameter is "1". The \textit{K} parameter is used in GenSSS model~\cite{Kurt2020MAM, Kurt2021TVC}, and it provides a control over the visual quality of GenSSS representation. It also affects the rendering times of GenSSS model linearly. More details about \textit{K} parameter can be found in Kurt~\cite{Kurt2020MAM, Kurt2021TVC}.  

\begin{figure*}[t]
\centering     
\footnotesize
$$\begin{tabular}{c@{\hspace{0.005\linewidth}}c@{\hspace{0.005\linewidth}}c@{\hspace{0.005\linewidth}}c}
  \includegraphics[width=0.24\linewidth]{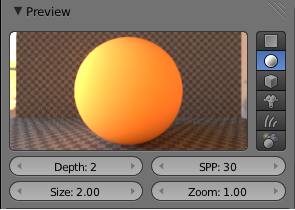}
  &\includegraphics[width=0.24\linewidth]{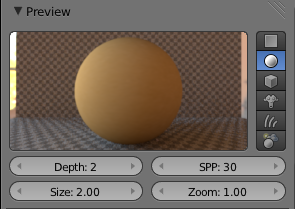}
  &\includegraphics[width=0.24\linewidth]{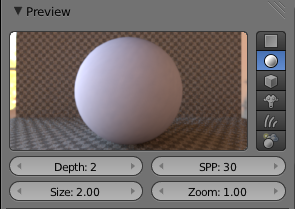}
  &\includegraphics[width=0.24\linewidth]{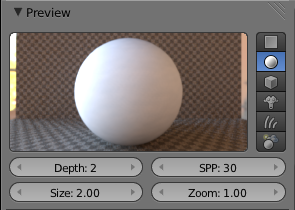}
    \\ \mbox{Yellow wax material}  &   \mbox{Jade material}  &   \mbox{Blue wax material} &   \mbox{Artificial stone material}\\
  \includegraphics[width=0.24\linewidth]{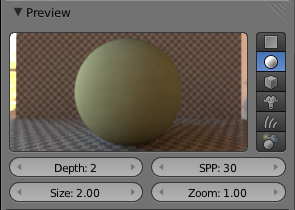}
  &\includegraphics[width=0.24\linewidth]{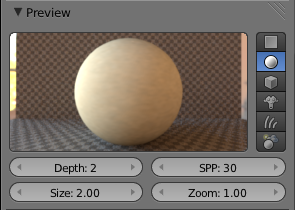}
  &\includegraphics[width=0.24\linewidth]{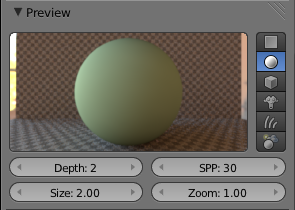}
  &\includegraphics[width=0.24\linewidth]{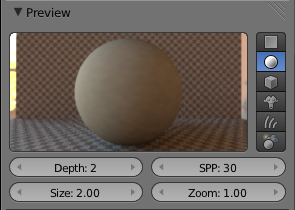}
  \\  \mbox{Chessboard ($4\times4$) material} & \mbox{Chessboard ($8\times8$) material}  &  \mbox{Marble (close up) material}  & \mbox{Densely veined marble material}
\end{tabular}$$
\vspace{-0.4cm}
\caption{\label{fig:homoprewrenders} A sphere was rendered at preview scene of our integration plugin (GenPluSSS) for various homogeneous translucent materials.}
\vspace{-0.4cm}
\end{figure*}

\begin{figure*}[t]
\centering     
\footnotesize
$$\begin{tabular}{c@{\hspace{0.005\linewidth}}c@{\hspace{0.005\linewidth}}c@{\hspace{0.005\linewidth}}c}
  \includegraphics[width=0.24\linewidth]{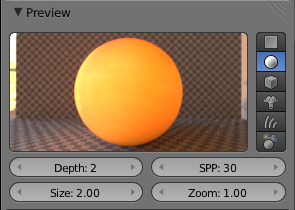}
  &\includegraphics[width=0.24\linewidth]{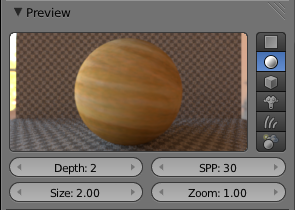}
  &\includegraphics[width=0.24\linewidth]{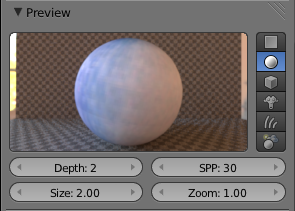}
  &\includegraphics[width=0.24\linewidth]{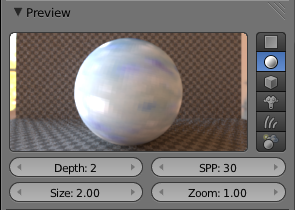}\\Yellow wax material  &  Jade material  &  Blue wax material &  Artificial stone material \\
  \includegraphics[width=0.24\linewidth]{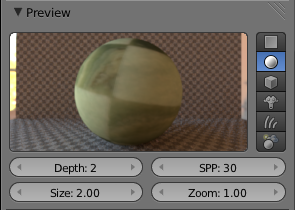}
  &\includegraphics[width=0.24\linewidth]{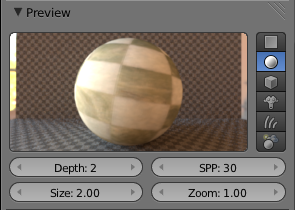}
  &\includegraphics[width=0.24\linewidth]{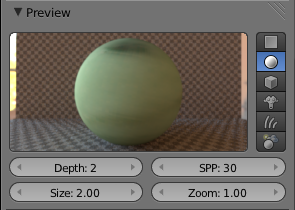}
  &\includegraphics[width=0.24\linewidth]{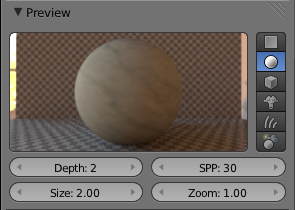}
  \\  Chessboard ($4\times4$) material & Chessboard ($8\times8$) material  & Marble (close up) material  & Densely veined marble material
\end{tabular}$$
\vspace{-0.4cm}
\caption{\label{fig:heteroprewrenders} A sphere was rendered at preview scene of our integration plugin (GenPluSSS) for various heterogeneous translucent materials. The value of \textit{K} parameter in GenSSS model was selected as 10.}
\vspace{-0.4cm}
\end{figure*}

Graphical User Interface (GUI) of our proposed integration plugin is shown in Figure~\ref{fig:selectscreen}. The representation of homogeneous and heterogeneous translucent materials is classified under the section of Mitsuba Subsurface Scattering, which is readily available in Mitsuba integration plugin~\cite{BlenderPlugin}. The details of the material type is chosen by using the GUI of the proposed plugin (see Figure~\ref{fig:blenderview} and Figure~\ref{fig:selectscreen}). As the Blender project~\cite{Blender} supports to model different types of materials and other details in the scene, the effect of the chosen operation is completed after the rendering operation.

\section{Experimental Results}
\begin{figure*}[t]
\centering     
\footnotesize
$$\begin{tabular}{c@{\hspace{0.005\linewidth}}c@{\hspace{0.005\linewidth}}c@{\hspace{0.005\linewidth}}c}
  \includegraphics[width=0.24\linewidth]{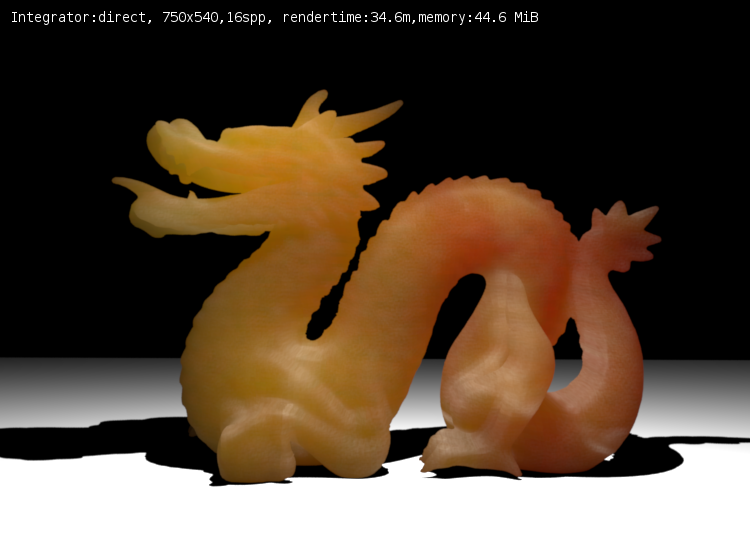}
  &\includegraphics[width=0.24\linewidth]{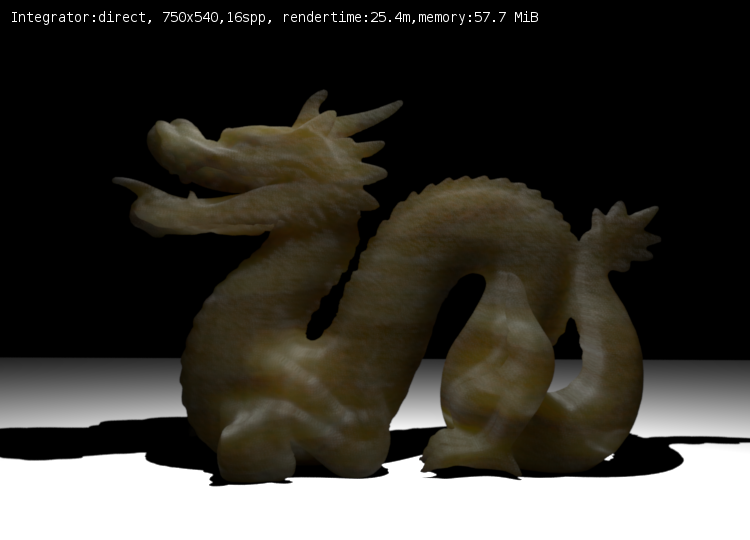}
  &\includegraphics[width=0.24\linewidth]{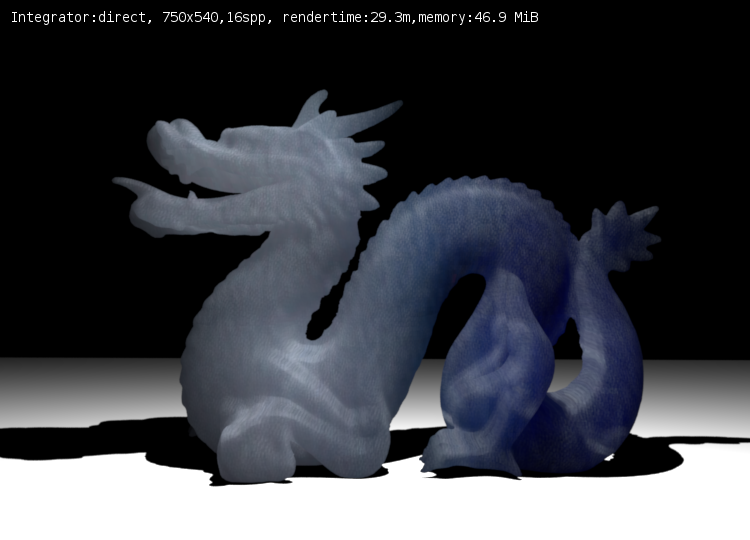}
  &\includegraphics[width=0.24\linewidth]{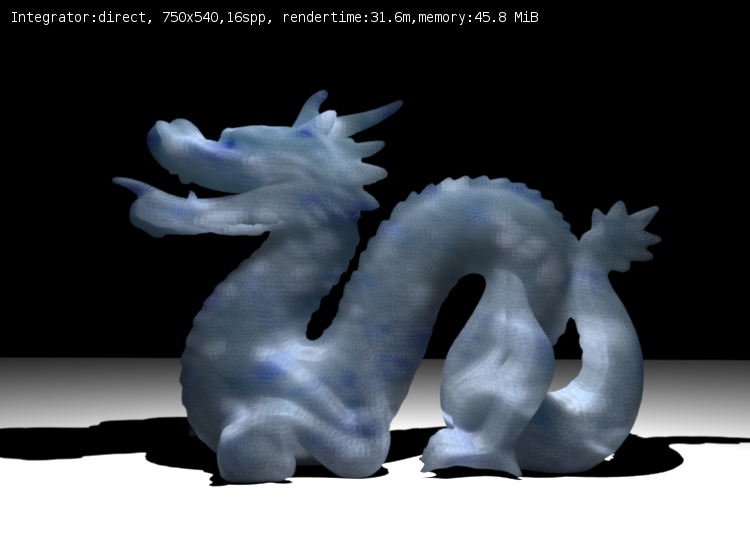}
  \vspace{-0.4cm}
  \\Yellow wax material  &  Jade material  &  Blue wax material &  Artificial stone material
  \\Data size: 46.8 MB &  Data size: 60.5 MB  &  Data size: 49.2 MB &  Data size: 48 MB
  \\Rendering time: 34.6 min. & Rendering time: 25.4 min. & Rendering time: 29.3 min. & Rendering time: 31.6 min. \\
  \includegraphics[width=0.24\linewidth]{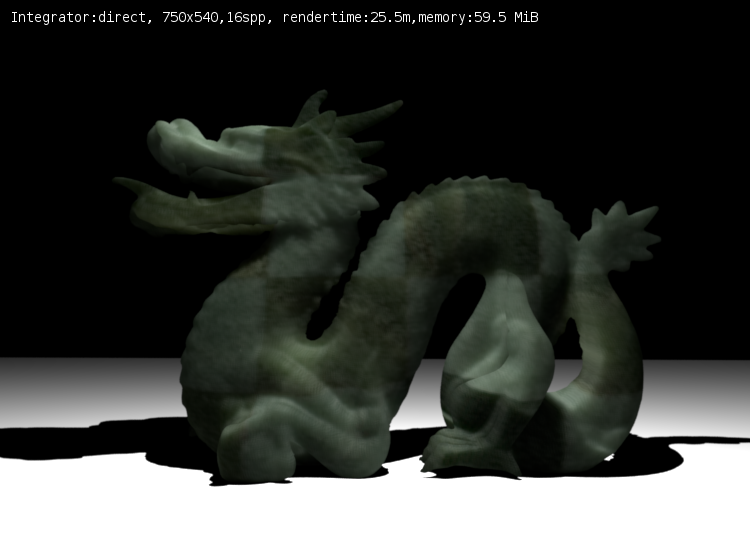}
  &\includegraphics[width=0.24\linewidth]{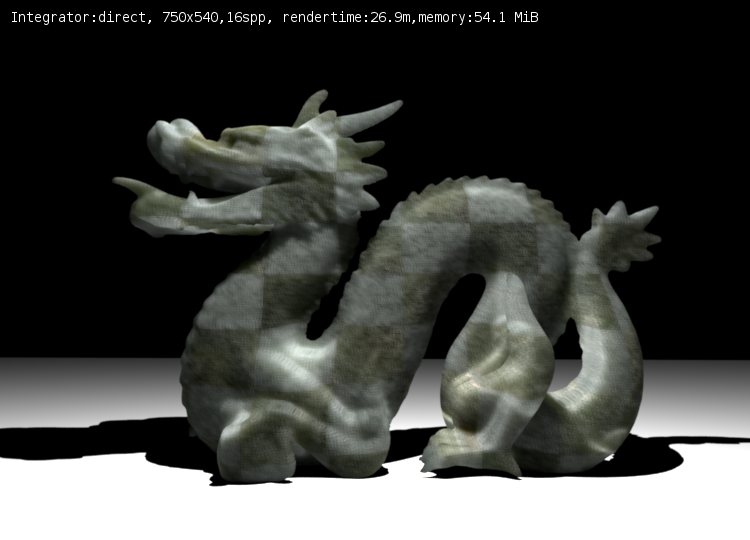}
  &\includegraphics[width=0.24\linewidth]{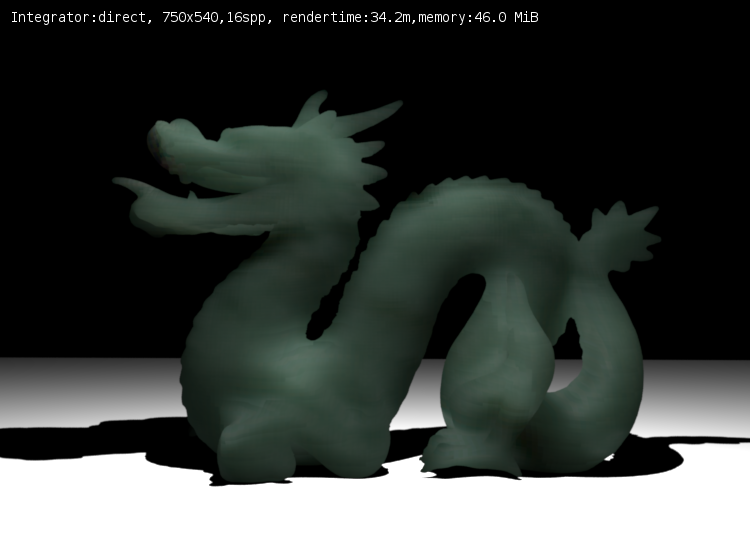}
  &\includegraphics[width=0.24\linewidth]{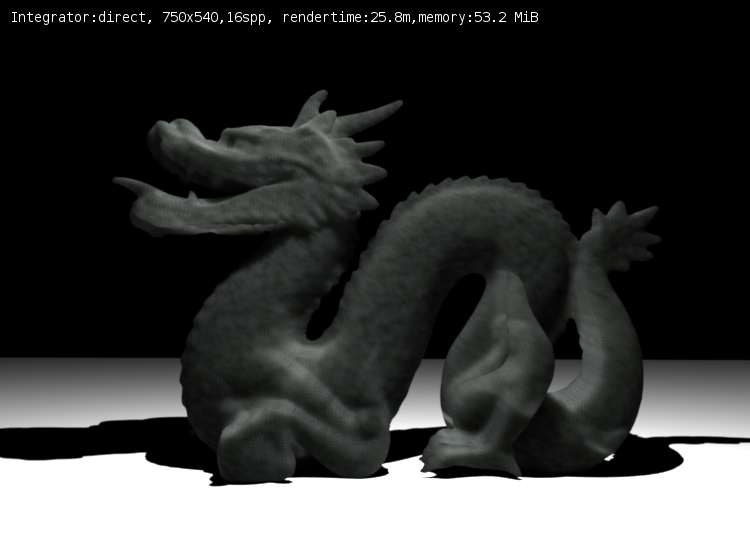}
  \vspace{-0.4cm}
  \\  Chessboard ($4\times4$) material & Chessboard ($8\times8$) material  & Marble (close up) material  & Densely veined marble material
  \\Data size: 62.4 MB &  Data size: 56.7 MB  &  Data size: 48.2 MB &  Data size: 55.8 MB
  \\Rendering time: 25.5 min. & Rendering time: 26.9 min. & Rendering time: 34.2 min. & Rendering time: 25.8 min.
\end{tabular}$$
\vspace{-0.4cm}
\caption{\label{fig:hetorenders} Various subsurface scattering representations for different heterogeneous translucent materials with their rendering times and required storage spaces. A dragon under spot lighting was rendered using our integration plugin (GenPluSSS). The value of \textit{K} parameter in GenSSS model was selected as 10.}
\vspace{-0.4cm}
\end{figure*}

\begin{figure*}[t]
\centering     
\footnotesize
$$\begin{array}{c@{\hspace{0.005\linewidth}}c@{\hspace{0.005\linewidth}}c@{\hspace{0.005\linewidth}}c}
  \includegraphics[width=0.24\linewidth]{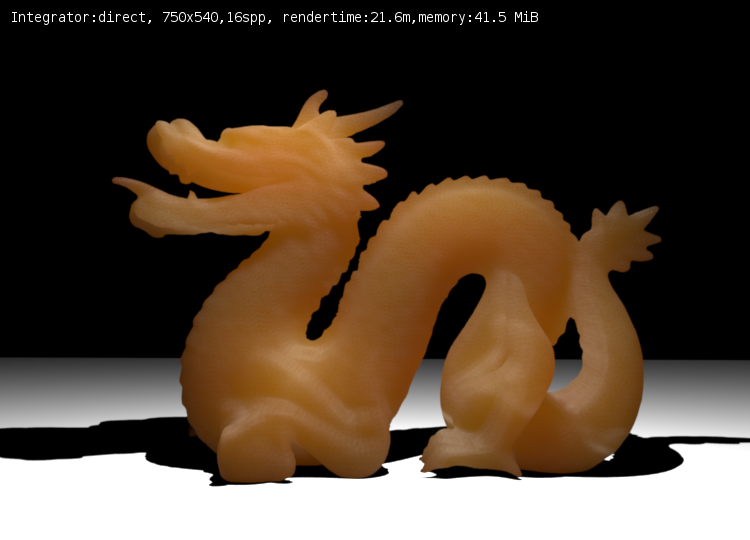}
  &\includegraphics[width=0.24\linewidth]{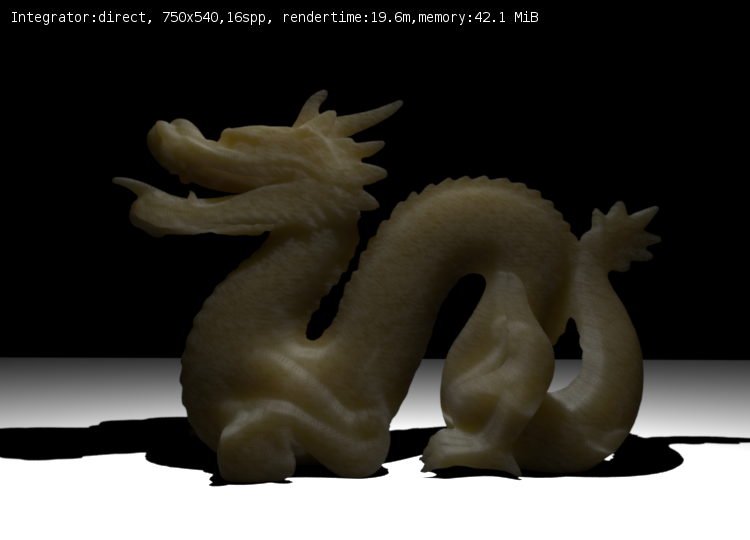}
  &\includegraphics[width=0.24\linewidth]{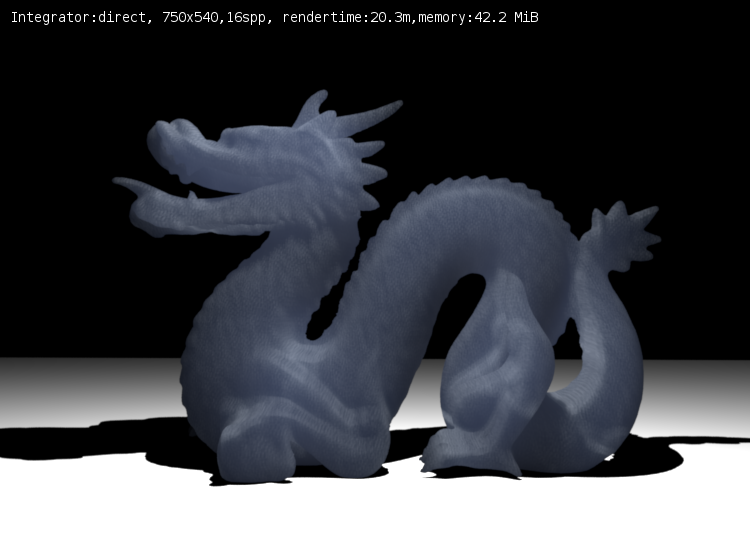}
  &\includegraphics[width=0.24\linewidth]{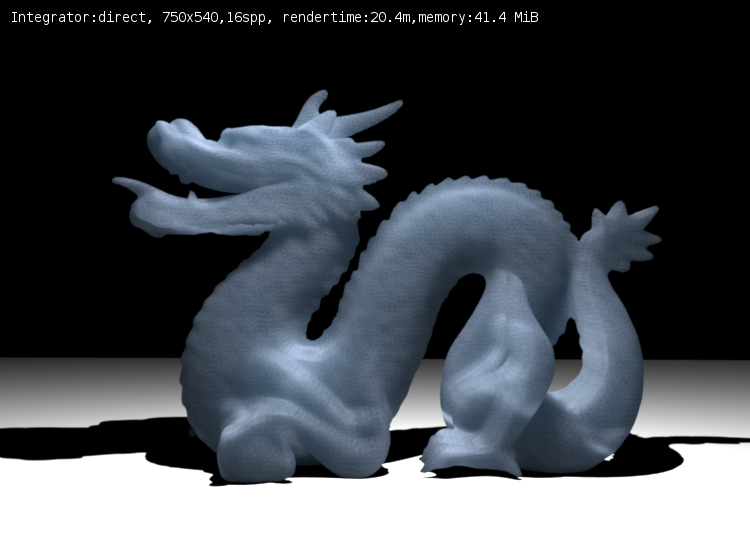}
  \vspace{-0.4cm}
  \\\mbox{Yellow wax material}  &   \mbox{Jade material}  &  \mbox{Blue wax material} &   \mbox{Artificial stone material} 
  \\\mbox{Data size: 43.5 MB} &   \mbox{Data size: 44.1 MB}  &   \mbox{Data size: 44.2 MB} &   \mbox{Data size: 43.4 MB} 
  \\\mbox{Rendering time: 21.6 min.} &  \mbox{Rendering time: 19.6 min.} &  \mbox{Rendering time: 20.3 min.} &  \mbox{Rendering time: 20.4 min.}\\
  \includegraphics[width=0.24\linewidth]{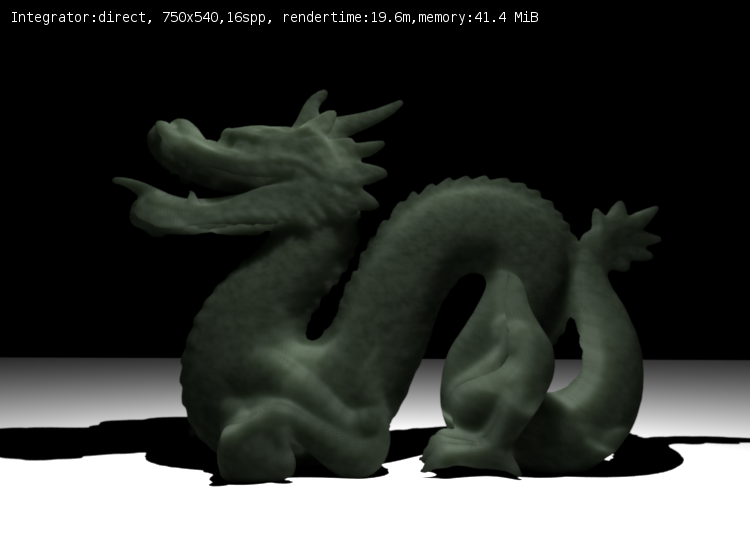}
  &\includegraphics[width=0.24\linewidth]{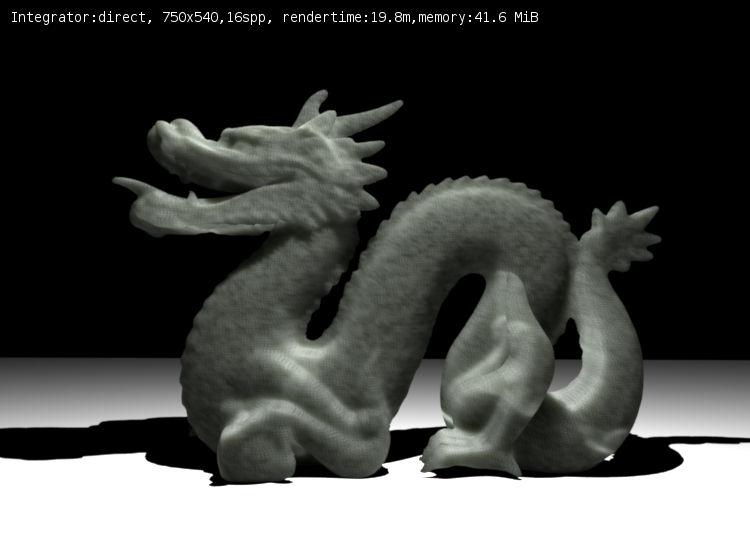}
  &\includegraphics[width=0.24\linewidth]{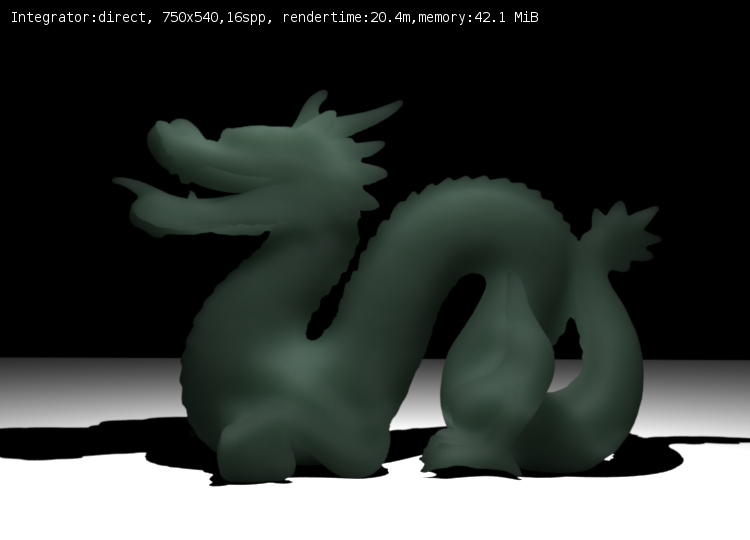}
  &\includegraphics[width=0.24\linewidth]{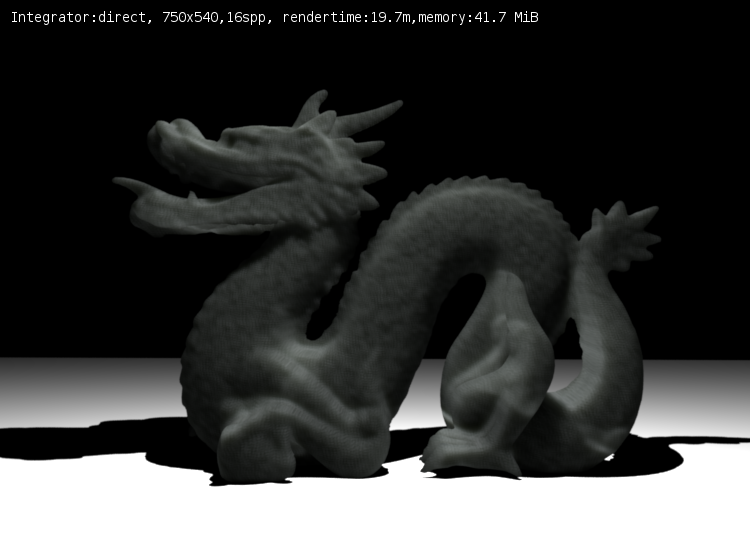}
  \vspace{-0.4cm}
  \\  \mbox{Chessboard ($4\times4$) material} & \mbox{Chessboard ($8\times8$) material}  & \mbox{Marble (close up) material}  & \mbox{Densely veined marble material}
  \\ \mbox{Data size: 43.4 MB} &  \mbox{Data size: 43.6 MB}  &  \mbox{Data size: 44.1 MB} &  \mbox{Data size: 43.7 MB}
  \\ \mbox{Rendering time: 19.6 min.} & \mbox{Rendering time: 19.8 min.} & \mbox{Rendering time: 20.4 min.} & \mbox{Rendering time: 19.7 min.}
\end{array}$$
\vspace{-0.4cm}
\caption{\label{fig:homorenders} Various subsurface scattering representations for different homogeneous translucent materials with their rendering times and required storage spaces. A dragon under spot lighting was rendered using our integration plugin (GenPluSSS). The value of \textit{K} parameter in GenSSS model is 1.}
\vspace{-0.4cm}
\end{figure*}

The proposed plugin was validated on various translucent materials, which were measured by Peers et al.~\cite{Peers06} and Song et al.~\cite{Song09}. Firstly, subsurface scattering measurements were modeled by the genetic optimization and SVD-based subsurface scattering representation (GenSSS). Results show that our proposed plugin provides homogeneous and heterogeneous subsurface scattering effects visually plausibly (see Figure~\ref{fig:hetorenders}, and Figure~\ref{fig:homorenders}). Also, two issues were reported in the text. These are the processing time that the plugin requires to complete the rendering process and the real-time storage requirements.

The materials chosen for the validation were applied on a dragon object. We checked whether the plugin could perform subsurface scattering effects on the objects as it does directly with Mitsuba renderer. We allowed to select "Material Type", "Material Name", "Parameter" in our proposed plugin (see Figure~\ref{fig:blenderview}, and Figure~\ref{fig:selectscreen}). The details of the scene were exported to the renderer successfully.


Figure~\ref{fig:homoprewrenders} illustrates the preview of homogeneous subsurface scattering effects and Figure~\ref{fig:heteroprewrenders} illustrates the preview of heterogeneous subsurface scattering effects on a sphere solid object. Both homogeneous subsurface scattering effects and heterogeneous subsurface scattering effects are demonstrated with yellow wax, jade, blue wax, artificial stone, chessboard ($4\times4$), chessboard ($8\times8$), marble (close up) and densely veined marble materials at preview scenes. In the preview scene in Figure~\ref{fig:heteroprewrenders}, the value of \textit{K} parameter in GenSSS model was selected as 10. A similar effect can be rendered on a different object. As it is seen in Figure~\ref{fig:hetorenders} and Figure~\ref{fig:homorenders}, our plugin (GenPluSSS) helps to render heterogeneous and homogeneous translucent materials correctly on a computer with i5-3210M processor with 6 GB RAM and NVIDIA GeForce GT 630M. Figure~\ref{fig:hetorenders} shows the various subsurface scattering representations, required storage spaces and real-time rendering times of using GenSSS with different heterogeneous translucent materials which was the selected value of \textit{K} parameter was '10'. Figure~\ref{fig:homorenders} illustrates the results of using GenSSS with different homogeneous translucent materials. 

\begin{figure}[t]
    \centering
    \includegraphics[width=0.94\linewidth]{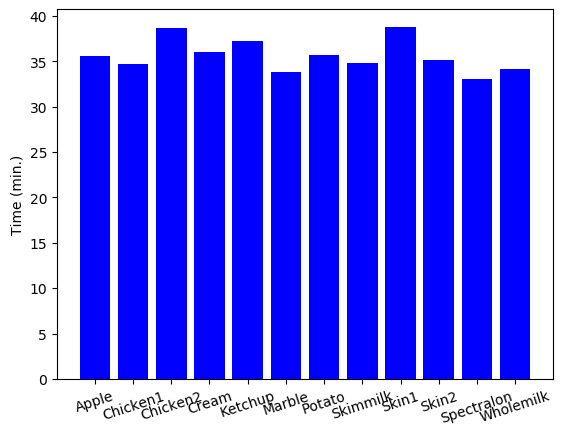}
    \vspace{-0.4cm}
    \caption{A bar graph that shows the rendering times of some homogeneous translucent materials that was presented by Jensen et al.~\cite{Jensen01}.}
    \label{fig:barGraphJensen}
    \vspace{-0.4cm}
\end{figure}

\begin{figure}[t]
    \centering
    \includegraphics[width=0.94\linewidth]{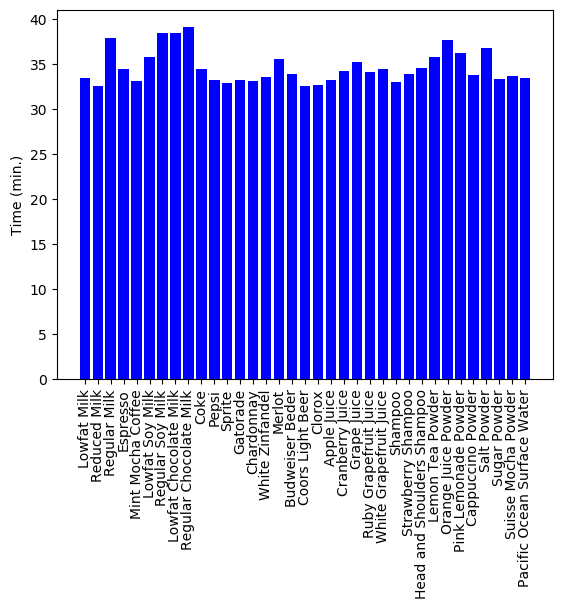}
    \vspace{-0.4cm}
    \caption{A bar graph that shows the rendering times of some homogeneous translucent materials that was presented by Narasimhan et al.~\cite{narasimhan2006acquiring}.}
    \label{fig:barGraphNarasimhan}
    \vspace{-0.4cm}
\end{figure}

\begin{figure}[t]
    \centering
    \includegraphics[width=0.94\linewidth]{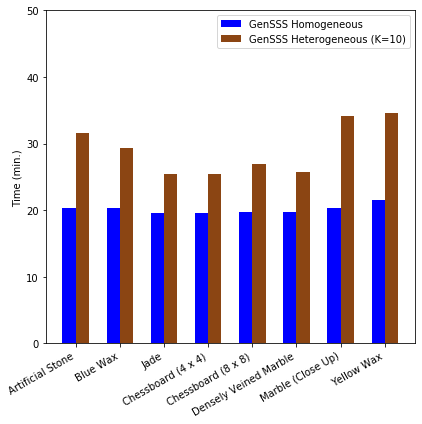}
    \vspace{-0.4cm}
    \caption{A bar graph that shows comparison of the rendering times of some homogeneous and heterogeneous translucent materials that was calculated with GenSSS method.}
    \label{fig:barGraphHomoHetero}
    \vspace{-0.4cm}
\end{figure}

\begin{figure}[t]
    \centering
    \includegraphics[width=0.94\linewidth]{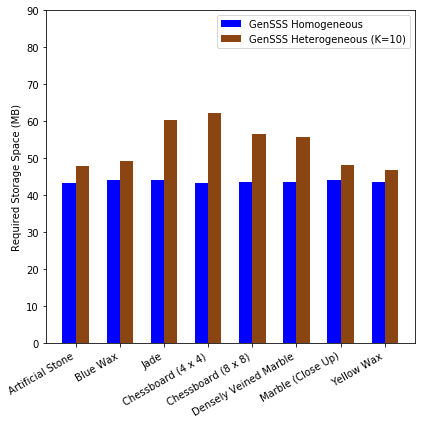}
    \vspace{-0.4cm}
    \caption{A bar graph that shows comparison of the required storage spaces for the rendering of some homogeneous and heterogeneous translucent materials that was calculated with GenSSS method.}
    \label{fig:bargraphStorageSpace}
    \vspace{-0.4cm}
\end{figure}

Figure~\ref{fig:barGraphJensen} illustrates the rendering times of 12 homogeneous translucent materials that were presented by Jensen et al.~\cite{Jensen01}, Figure~\ref{fig:barGraphNarasimhan} illustrates the rendering times of 35 homogeneous translucent materials that were presented by Narasimhan et al.~\cite{narasimhan2006acquiring} and Figure~\ref{fig:barGraphHomoHetero} illustrates the rendering times of homogeneous and heterogeneous translucent materials that were calculated with GenSSS method. Figure~\ref{fig:bargraphStorageSpace} illustrates required storage spaces of GenSSS representation for different homogeneous and heterogeneous translucent materials. The value of \textit{K} parameter for heterogeneous translucent materials was selected as 10. However, the value of \textit{K} parameter for homogeneous translucent materials was 1 as default. When high \textit{K} values are preferred, the rendering times and required storage space increase. Therefore, with GenSSS model, the rendering times and required storage spaces of heterogeneous translucent materials are higher than homogeneous translucent materials, as it seen at Figure~\ref{fig:barGraphHomoHetero} and Figure~\ref{fig:bargraphStorageSpace}.

The average real-time rendering times and the average data sizes of homogeneous and heterogeneous translucent materials in GenSSS model were calculated to compare with other studies. With these data, the average real-time rendering time of homogeneous translucent materials was found to be 20.17 minutes and the output needed an average of 43.75 MB storage. The average rendering time of heterogeneous translucent materials was found to be 29.16 minutes and the average data size was 53.45 MB. Using the SVD method in the plugin that provided by \"{O}nel et al.~\cite{Onel2014EG,Onel2019PL}, the real-time rendering time of the heterogeneous chessboard ($4\times4$) material was found to be 25.1 minutes and the rendering time of the heterogeneous chessboard ($8\times8$) material was found to be 26.4 minutes. The required storage spaces were 63.54 MB and 55.89 MB for the chessboard ($4\times4$), and the chessboard ($8\times8$), respectively. Hence, the average rendering time of these materials was 25.75 minutes and the average data size was 59.69 MB. In the GenSSS method, the transformation is applied in addition to the SVD method and this situation causes the rendering time to increase, when it's compared to the SVD method~\cite{Onel2014EG}. 

In the exporter plugin that was developed by Styperek and Juh\'e~\cite{BlenderPlugin}, the real-time rendering times of 12 homogeneous translucent materials that were measured and presented by Jensen et al.~\cite{Jensen01} with the dipole approximation were tested (see Figure~\ref{fig:barGraphJensen}). These measurements were made in such a way that the sample numbers of the homogeneous translucent materials to which the dipole approach was applied and the sample numbers of the homogeneous translucent materials in the GenSSS method were equal in order to make a fair comparison. As a result of the measurements, the average rendering time of these 12 homogeneous translucent materials was calculated as 35.6 minutes despite of the average rendering time of 8 homogeneous translucent materials (see Figure~\ref{fig:barGraphHomoHetero}) calculated by the GenSSS method was only 20.17 minutes. In addition to the average rendering time, the average data size was 43.93 MB. Also, 35 homogeneous translucent materials which were measured by Narasimhan et al.~\cite{narasimhan2006acquiring} were tested with the exporter plugin (see Figure~\ref{fig:barGraphNarasimhan}) and the average rendering time of these materials was found as 34.6 minutes. An average of 43.82 MB of storage space was required for these materials. As it's seen from these results, results are obtained faster in GenSSS model for rendering homogeneous translucent materials. In GenSSS model, \textit{range} parameter is as important as \textit{K} parameter on the rendering times. Genetic Algorithm helps for finding the optimum range of applied transformations, which also optimize the rendering times of GenSSS model. In addition this, when the value of \textit{K} parameter is 1, it's easy to compute of representation with GenSSS model. Generally, analytical subsurface scattering models require less storage than data-driven subsurface scattering models. Because data-driven subsurface scattering presentations require more data. However, in our case, GenSSS model needed less storage space, because the value of \textit{K} parameter is set to 1 while representing homogeneous materials.


\section{Conclusion}
In this paper, we introduced our integration plugin (GenPluSSS) for rendering of various homogeneous and heterogeneous translucent materials. Our plugin (GenPluSSS) is based on GenSSS model which proposed by Kurt~\cite{Kurt2020MAM,Kurt2021TVC} and this plugin made possible to render visually acceptable scenes. We chose the Blender~\cite{Blender} 3D modeling tool to prepare our integration plugin (GenPluSSS). We also showed the homogeneous and heterogeneous subsurface scattering effects which we rendered on the sphere object in the preview scene in the Blender 3D modeling tool. In addition to these, we also made comparisons between our plugin (GenPluSSS) and the other integration plugins~\cite{BlenderPlugin,Onel2014EG} for the average real-time rendering times and the average data sizes. As a result of these, it has been observed that our plugin (GenPluSSS) enables a fast result for render times in homogeneous translucent materials. It has also been shown that it can quite accurately represent subsurface scattering effects in homogeneous and heterogeneous translucent materials.

\section{Future Works}
In this study, various homogeneous and heterogeneous translucent materials were rendered using the GenSSS model in Blender~\cite{Blender} 3D modeling tool. For the real-time rendering, Blender 3D modeling tool which is an open source development platform and which uses Python~\cite{Python} as a scripting language was chosen; because Mitsuba renderer~\cite{Mitsuba} has been written by C++ programming language and C++ functions can be easily called using the Python programming language. 

As a future work, this plugin (GenPluSSS) can be developed to use at different platforms. Thus, we are interested in providing for the representation of homogeneous and heterogeneous subsurface scattering effects at different 3D modeling tools. We are also interested in representing BRDFs~\cite{Ozturk2006EGUK,Kurt2007MScThesis,Ozturk2008CG,Kurt2008SIGGRAPHCG,Kurt2009SIGGRAPHCG,Kurt2010SIGGRAPHCG,Ozturk2010GraphiCon,Ozturk2010CGF,Bigili2011CGF,Bilgili2012SCCG,Tongbuasirilai2017ICCVW,Kurt2019DEU,Akleman2024arXiv}, BSDFs~\cite{WKB12,Ward2014MAM,Kurt2014WLRS,Kurt2016SIGGRAPH,Kurt2017MAM,Kurt2018DEU}, and multi-layered materials~\cite{WKB12,Kurt2016SIGGRAPH,Mir2022DEU} at various 3D modeling tools.


\bibliographystyle{IEEEtran}
\bibliography{arXiv24_GenPluSSS_References_V1}

\vspace{12pt}
\color{red}
\end{document}